\pdfoutput=1 
\documentclass[english,nohyper,a4paper]{tufte-handout}
\usepackage{amsmath,amssymb,amsthm}
\usepackage{mathpple}

\usepackage[latin9]{inputenc}
\usepackage[T1]{fontenc}
\usepackage{babel}
\usepackage{color}
\usepackage{wasysym}   
\usepackage{marvosym}  
\usepackage{graphicx}
\usepackage{microtype}
\def\adx#1:#2\par{\par\halign{\hskip #1##\hfill\cr #2}\par}

\def\gcm3{g~cm$^{-3}$}

\def\mesa{\texttt{MESA}}
\def\rsp{\texttt{RSP}}

\def\teff{T_{\rm eff}}
\def\mast{M_\ast}

\def\msol{M_\odot}
\def\lsol{L_\odot}

\def\rast{R_\ast}

\def\llsol{L/L_\odot}

\def\H1{^1\mathrm{H}}
\def\He3{^3\mathrm{He}}
\def\He4{^4\mathrm{He}}
\def\C12{^{12}\mathrm{C}}
\def\N14{^{14}\mathrm{N}}
\def\O16{^{16}\mathrm{O}}

\def\diff{{\mathrm d}}

\def\unity{ \hbox{1\kern-.23em l} }
\def\zero{ \hbox{0\kern-.23em |} }
\def\field{ \hbox{I\kern-.23em K} }

\def\braket #1.#2.{\langle #1 \vert #2 \rangle}

\def\O{\mathcal{O}}

\newcommand{\lyxaddress}[1]{ \vspace{1.4em} \par {\raggedright #1 \vspace{1.4em}
\noindent\par} }

\usepackage{amsmath}
\usepackage{booktabs}
\title{Overtone Pulsations Revisited} \author{Alfred Gautschy}
\date{}

\begin{document} \maketitle  


\lyxaddress{CBmA, 4410 Liestal, Switzerland}

\begin{abstract} \noindent The shape of light cures of
  fundamental-mode and of first-overtone pulsators, as observed
  in RR~Lyrae variables and Cepheids, differ characteristically.
  The stellar physical origin of the morphological differences is
  not well documented and the topic seems not to be part of the elementary
  curriculum of students of stellar variability. To ameliorate the
  situation, this exposition analyzes hydrodynamical simulations of radial
  pulsations computed with the newly available numerical 
  instrument \rsp\,in \mesa. The stellar
  physical processes that affect the light curves are identified and
  contrasted to the explanation based on experiments with one-zone models.
  The encounter of first-overtone pulsators sporting light curves that mimic those of 
  fundamental-mode variables serves as a warning that light-curve morphology
  alone is not a reliable path to correctly classify pulsating variables;
  the exposition closes with a short discussion of constraints these modes might
  impose on understanding the origin of anomalous Cepheids.  
\end{abstract}

\section{Introduction}
Temporal variability of celestial objects was and still is mainly detected
by measuring their brightness changes. The time-scale of the variability
together with the form of the temporal brightness change~--~the light 
curve~--~serve as the basic initial criteria to classify an object's 
variability. Reliable classifications, in particular ones which can be quantified 
and attributed automatically by software tools is gaining importance
in the age of petabyte data-generating sky surveys.

In the study of pulsating variable stars, Fourier decomposition of 
light curves has proved useful \citep{schaltenbrand_light_1971}. 
Amplitude ratios and phase differences of low-order components of invoked 
Fourier series \citep[e.g.][]{simon_structural_1981} 
are used to capture pertinent features of the light curves aiming at a 
robust classification of variable stars. For a long time stellar-pulsation
researchers even hoped that amplitude ratios and phase differences can be 
correlated with physical processes that operate in the interior of pulsating 
variables. Success was, however, partial at best.     

\citet{stellingwerf_fourier_1987} introduced a coarse-grained
light-curve characterization via the concepts of 
\emph{Skewness }(Sk) and \emph{Acuteness }(Ac).\sidenote[][-1.5  cm]{
     Sk~$\doteq  1/\Delta\phi_{\mathrm{rb}} - 1$
	with $\Delta\phi_{\mathrm{rb}}$ measuring the fraction of the 
	pulsation period taken up by the rising branch of the light curve. 
	Ac~$\doteq  1/\Delta\phi_{\mathrm{fw}} - 1$ where $\Delta\phi_{\mathrm{fw}}$
	measures the time (in units of the star's pulsation period) during which the	
	variable star stays above its mean brightness. Therefore, Ac is a 
	\emph{full-width-at-half-maximum} measure of the light curve.}
Based on simple nonlinear modeling behavior, the quantity Sk was observed 
to correlate with the nonlinearity of the pulsation velocity
whereas Ac was claimed to depend on the light amplitude at the base of 
the pulsation driving region. Since stellar pulsations are examples of
thermo-mechanical oscillators it is likely
that Ac and Sk are not independent of each other. 
Both quantifiers have been shown to correlate with
low-order Fourier components  but they are less sensitive measures of light-curve forms
than Fourier phases and amplitude ratios \citep{stellingwerf_fourier_1987}.
Irrespective of any physical interpretation of Sk and Ac, here they serve as useful 
morphological quantifiers in our endeavor to understand the 
\emph{Grundform}\sidenote[][-0cm]{in the sense of the \emph{uncarved block} that 
captures the constitutive shape already but does not yet reproduce elaborate 
overlying details}
of light curves of F- and 1O-mode pulsators.

\begin{marginfigure} 
	\begin{center}
	\includegraphics[width=0.97\textwidth]{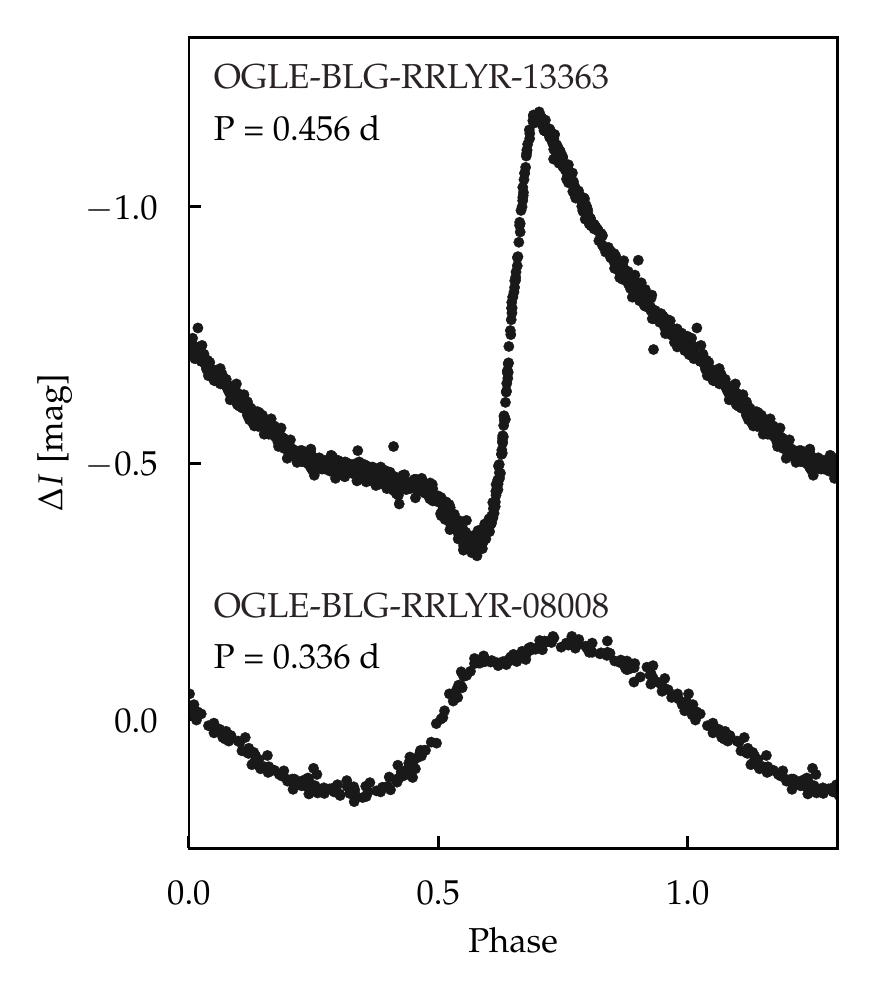} 
	\end{center}
     \caption{Light curves in the photometric I band of RR Lyrae variables 
              observed during the OGLE sky monitoring program.
              Top: RRab with Sk$= 5.9$, Ac$ = 1.8$.
              Bottom: RRc with Sk$ = 1.4$, Ac$ = 0.8$.
             } \label{fig:RRLyrObs}
\end{marginfigure}

Even before the RR~Lyrae variable stars were baptized as such and even 
before their having been recognized as a class they were abundantly 
observed globular clusters: In a study of the variable stars in $\omega$~Cen, 
\citet{bailey_discussion_1902} realized that the
\emph{cluster variables}, as the RR~Lyr variables were referred to then, 
come in more than one flavor. Guided by their periods, light-variation amplitude, 
and the form of the light curves, Bailey subdivided the cluster variables
into two longer-period, higher-amplitude groups with asymmetric light curves, 
which he termed RRa  
($ 0.5 \lesssim P/{\mathrm d} \lesssim 0.66$), cf. upper light curve 
in Fig.~\ref{fig:RRLyrObs}, RRb 
($ 0.66 \lesssim P/{\mathrm d} \lesssim 0.9$ and light curves somewhat
less acute than those of RRa variables) stars, 
and shorter-period ($0.3\lesssim P/{\mathrm d} \lesssim 0.5$), 
lower-amplitude variables with almost sinusoidal light curve shapes, 
were referred to as the RRc subclass, cf. lower light curve
in Fig.~\ref{fig:RRLyrObs}. 
When \citet{Schwarzschild1940} analyzed cluster-variable data of M3 
he speculated that RRc variables are first-overtone (1O) and 
the RRa and the RRb variables (conveniently fused to RRab) 
fundamental-mode (F) radially pulsating stars. This conjecture 
was later confirmed: \citet{Baker1965} presented results of linear stability analyses  
where 1O and F modes were excited simultaneously over essentially the full width
of the instability strip. In the same conference proceedings, \citet{Christy1965}
discussed his early nonlinear hydrodynamical computations of RRab and RRc variables. 
The light-curve shapes of the RRab models agreed well with observations.
The quality of the bolometric RRc light curves was, however, low; 
despite being less skew than F-mode light curves 
they lacked the low acuteness and the smoothness that characterize 
RRc variables. The situation improved markedly only toward the end 
of the last millennium \citep[e.g.][and references therein]{Feuchtinger1999}.

Despite some early successes to model hydrodynamically RR~Lyr variables pulsating 
in their F- and 1O-mode, respectively, a solid explanation of their 
characteristically differing  light curves was lacking.
\citet{stellingwerf_overtone_1987} resorted to the concept of 
one-zone models (OZM) in which the most important physics 
is boiled into parameters whose impact can be
easily studied with simplified nonlinear differential equations. Emphasizing the
effect of the spatial node of 1O modes on the luminosity that enters the pulsating 
"zone" of the OZM led  \citet{stellingwerf_overtone_1987} to conclude that
it is the phase reversal, relative to the luminosity variation inside the
pulsating shell, of the energy flux entering the pulsation zone
from the sub-node interior that effects the sinusoidal light curves of RRc stars. 
Taking the idea one mode order further, Stellingwerf et al. even speculated on how
the  light curves of RRe variables, i.e. RR~Lyr~--~type variables pulsating 
in their second overtone mode, should look like.   

Hydrodynamical modeling of radially pulsating stars has recently become more
widely accessible \citep{paxton2019}, the CPU power of even modest 
computers together with significantly broadened visualization possibilities 
led to this exposition that attempts to eventually check if the conclusions 
put forth in \citet{stellingwerf_overtone_1987} can be recovered in full 
hydrodynamical simulations. In a first step, RR~Lyr models are addressed. 
The analyses are carried over to Cepheids in whose short-period domain 
F and 1O modes can again be excited simultaneously so that different 
pulsation modes can be studied in the same envelope models, so that structural 
differences are eliminated that might affect pulsation modes when
models with different stellar parameters need to be invoked. 
This exposition also addresses the light curves of second-overtone pulsations, 
at least in the case of Cepheid models and finally discusses the 
surprising sighting of an 1O pulsator that masquerades as an F-mode
and its consequences. 

 
\section{RR Lyrae variables}

Using the \rsp\, \citep{Smolec2008} instrument of \mesa ,
a representative RR~Lyrae variable envelope
model was computed: The adopted stellar parameters were $0.65\,\msol$ at $45\,\lsol$
and $\teff = 6900$~K with a chemically homogeneous composition of 
$X=0.75, Z=0.0014$. The inner boundary of the envelope, where the pulsation velocity
is assumed to vanish and where a constant energy flux enters from the stellar interior,
was set at $2\times10^6$~K. The envelope was subdivided into $230$ zones, with $80$
zones located at temperatures below $11\cdot 10^3$~K in the hydrostatic starting model.

\begin{figure} 
	\begin{center}
	\includegraphics[width=0.9\textwidth]{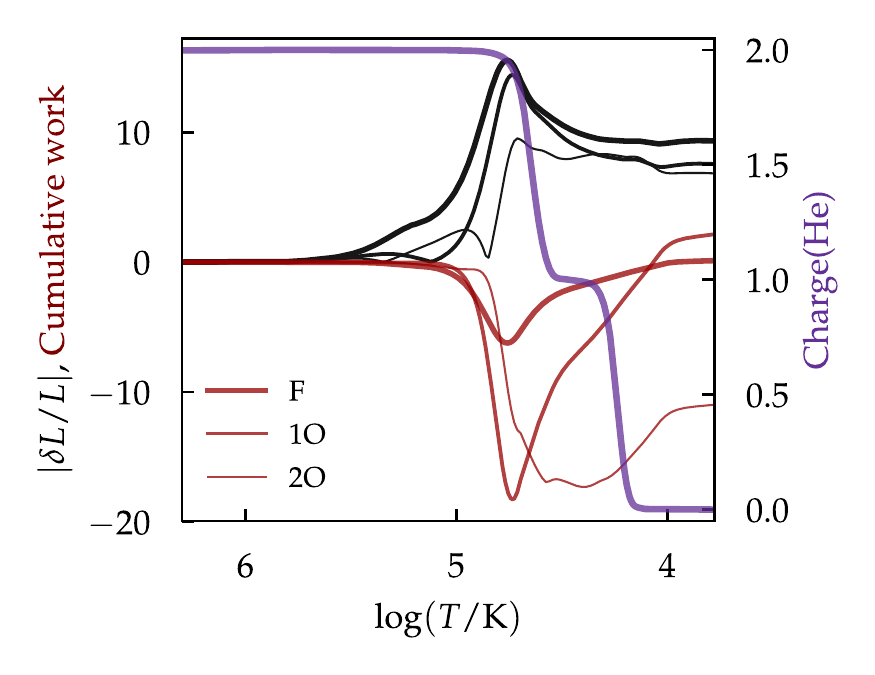} 
	\end{center}
     \caption{LNA results of the lowest three radial pulsation modes (coded
     	      by line thickness)
              of the RR~Lyrae model. Temperature measures the depth
              in the model's envelope.
              The relative luminosity perturbation is plotted with black lines.
              The cumulative work done by the envelope is coded in red. 
              For orientation, the indigo line traces the average charge 
              of helium atoms in the envelope.  
             } \label{fig:RRLyrLINA}
\end{figure}
Linear nonadiabatic (LNA) stability analysis of the hydrostatic initial model
revealed the RRab-like fundamental (F) mode  \emph{and }the RRc-like first overtone (1O)
mode to be both unstable. This allowed for the study of saturated nonlinear 
F and O1 pulsations of the same envelope model. 
Figure~\ref{fig:RRLyrLINA} displays pertinent quantities from the LNA 
analysis of the RR Lyr~--~type model: The black lines trace the spatial 
variation of the magnitude of the relative luminosity perturbation 
eigenfunctions $\vert \delta L/L \vert$ of the lowest three radial orders.
The legend in the lower left corner of the plot specifies the meaning of the
line thickness used for the eigendata.
The red lines trace the spatial structure of the cumulative work done by the respective
pulsation modes when integrating from the base of the envelope. A positive
value at the outer (right) boundary means that the respective mode is excited
(as it applies for F and 1O modes of the model shown in Fig.~\ref{fig:RRLyrLINA}). 
The remaining thick indigo-colored line measures the average charge of 
the helium atoms in the envelope. 
The rise of the average He-charge from 0 to 1 at about $ 4.15 < \log T < 4.4 $
traces the first partial ionization zone (PIZ) of helium (HeI). 
At slightly lower temperatures lies the PIZ of hydrogen. 
The HeII PIZ, with the average charge of helium changing from 1 to
2, spans the region of $ 4.55 < \log T < 4.8$.

\begin{figure*}
\label{fig:RRLyr_maps}
\centering
\includegraphics[width=0.48\linewidth]{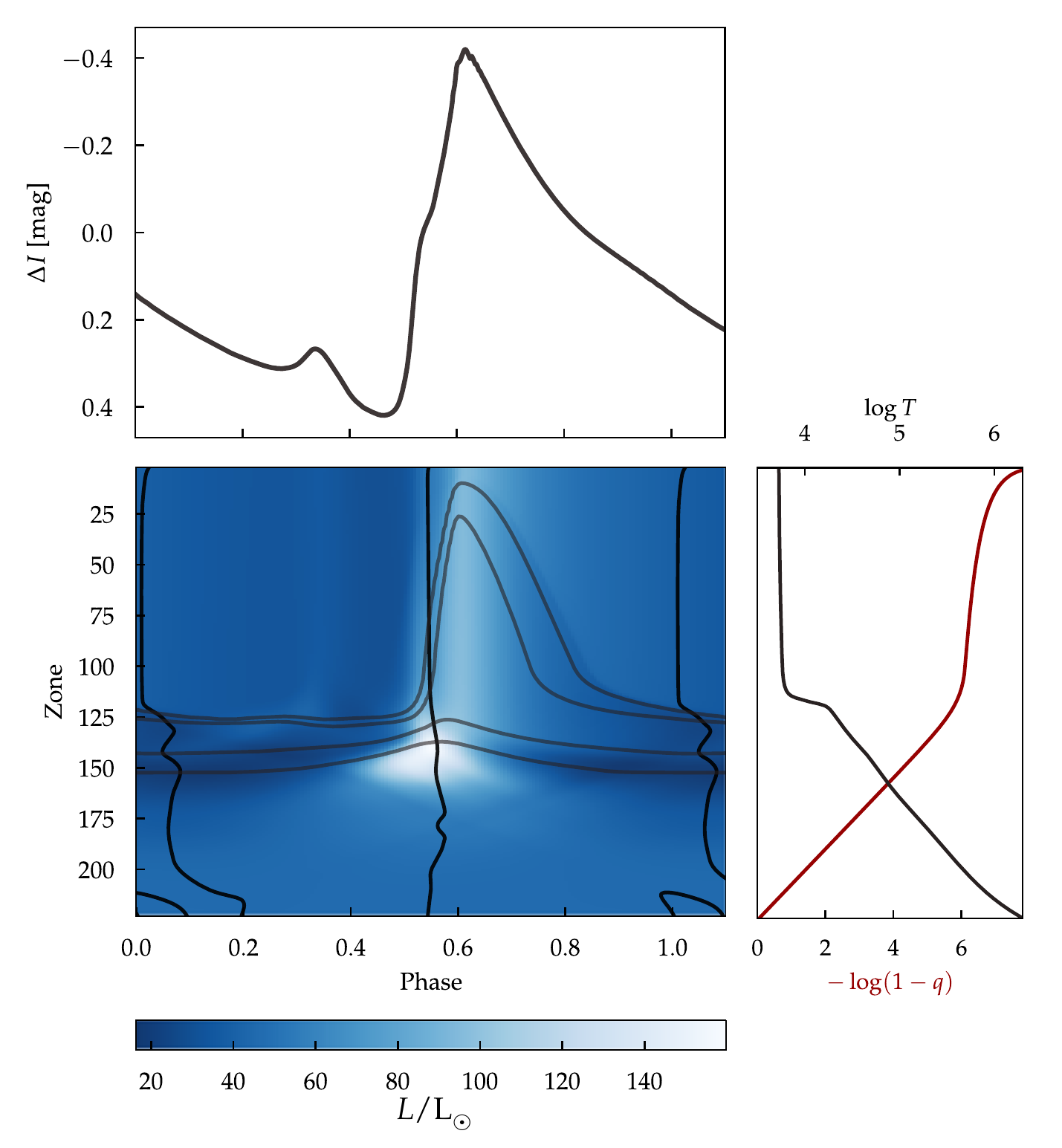}
\hspace{0.3cm}
\includegraphics[width=0.48\linewidth]{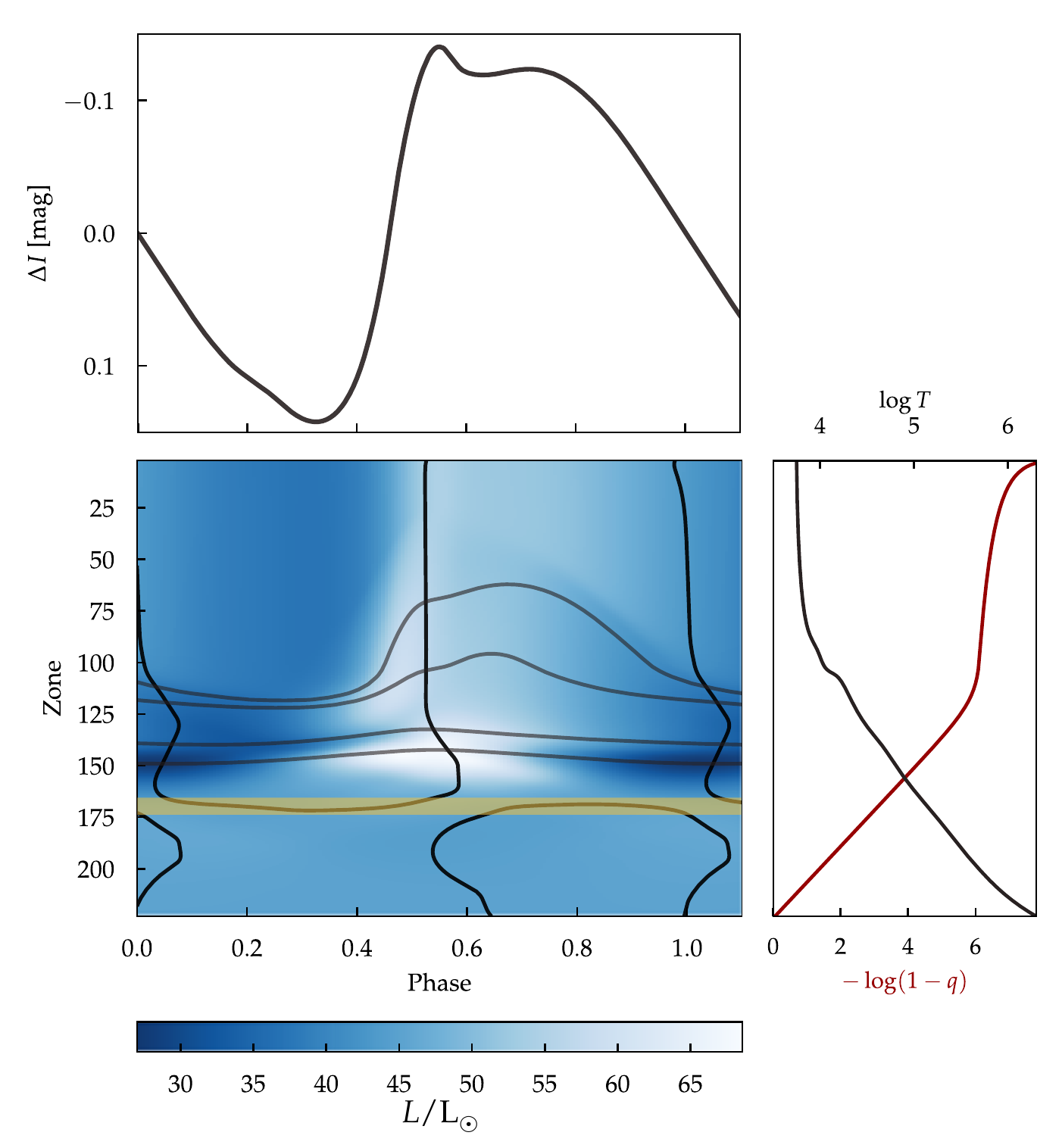}
\caption{Triptycha composed of two pulsation panels on top of each other and 
         a structure panel on the bottom right for the RRab (left, $P=0.4560$~d) 
         and the RRc (right, $P=0.3361$~d) model.
        }
         
\end{figure*}
The particular nonlinear mode aimed at in the direct integration of the
nonlinear hydro equations can frequently be selected
initially by imposing the appropriate LNA velocity eigenfunction as the perturbation of
the static start model 
\citep[cf. \mesa\, instrument paper][e.g. their Sect.~2.4.6 and references therein]{paxton2019}. Both, F and 1O modes
could be successfully followed into their respective limit cycles. 
Selected behavior of radial pulsations of the RRab- and RRc-type pulsations that
have reached their respective limit cycles is depicted in  Fig.~\ref{fig:RRLyr_maps}
Both triptycha are composed in the same way: The central frame is the color 
coded spatio-temporal bolometric luminosity variation over $1.2$ pulsation cycles;
phase~$ = 0$ is set at the epoch of maximum photospheric radius. 
The simulations' zone number serves as the spatial coordinate. The counting
starts at the stellar surface. Because the \rsp\,instrument 
is a lagrangian hydro code, zone numbers 
map onto a temporally invariant fixed-mass gridding. The color coding of the 
luminosity variation, in units of solar luminosity, is resolved a the bottom of
the frame. The thick black lines trace the contour of $v_{\mathrm{puls}}=0$. 
Both triptycha  show that the phases of maximum and minimum radius shift as 
a function of depth in the envelope. The considerable shifts close to the inner
boundary of the envelopes are likely affected by interpolation uncertainties at the
very small pulsation amplitudes in the deep interior.
The colormesh plot of the RRc-like pulsation in the right-hand triptychon
of Fig.~\ref{fig:RRLyr_maps} was augmented with a transparent yellow line tracing the
essentially horizontal parts of $v_{\mathrm{puls}}=0$ loci. 
This spatially invariant line approximates well the position of the 1O 
pulsation mode's  \emph{node} (which is also seen at about $\log T = 5$ 
in the luminosity perturbation of the LNA analysis shown in Fig.~\ref{fig:RRLyrLINA}).
Four weaker grey lines on top of the colormesh plots trace the mass depth
of the two helium PIZs; the lines trace the average He charge of 
$1.9, 1.1$ (HeII PIZ) and higher up in the envelope $0.9, 0.1$ (HeI PIZ), 
respectively. 

To get an impression of what the zone numbers mean physically, 
the plot panels in the lower right of the triptycha quantify the 
temporally invariant mass stratification 
(with $q\doteq m/\mast$ ) in red and the temperature stratification  
at phase~$= 0$ (black line). The emergent light curve in the photometric 
$I$-band of the saturated pulsation mode is plotted on top of the color frame, 
both panels span $1.2$ pulsation 
cycles.\sidenote{The $I$-band magnitudes were computed requesting 
\texttt{log\_lum\_band bb\_I} in \mesa's \texttt{history\_columns.list}; the "\texttt{bb}"
option assumes the stellar radiation field to be a blackbody with the model's 
effective temperature to compute the bolometric correction.
}

Most evidently, both colormesh frames of Fig.~\ref{fig:RRLyr_maps} show that 
the global maximum luminosity is not radiated from the stellar surface but is 
attained deeper in the stellar envelope. This property is, of course, also
recovered in the LNA analyses (cf. Fig.~\ref{fig:RRLyrLINA}). 
The pile-up of the star's energy flux happens in layers where helium goes from 
twice to once ionized. Figure~\ref{fig:RRLyr_maps} illustrates that also the 
darkest patches (lowest luminosities) are encountered at about the hotter 
end the HeII PIZ. 
This observation of the luminosity behavior applies to both, 
the RRab and the RRc models. In contrast to the F mode, the 1O pulsator 
experiences, however, a comparatively longer high-luminosity phase. 
The computations show that the luminosity blocking effect that takes 
place at the base of the HeII PIZ broadly shapes, albeit somewhat smoothed out, 
the emergent surface luminosity variation. Hence, what happens at the base
of the HeII PIZ governs the Grundform of the observable
light curve, and it is the quantity Ac, which is affected in particular. 

\begin{figure}
\label{fig:RRLyr_LandR}
\centering
\includegraphics[width=0.48\linewidth]{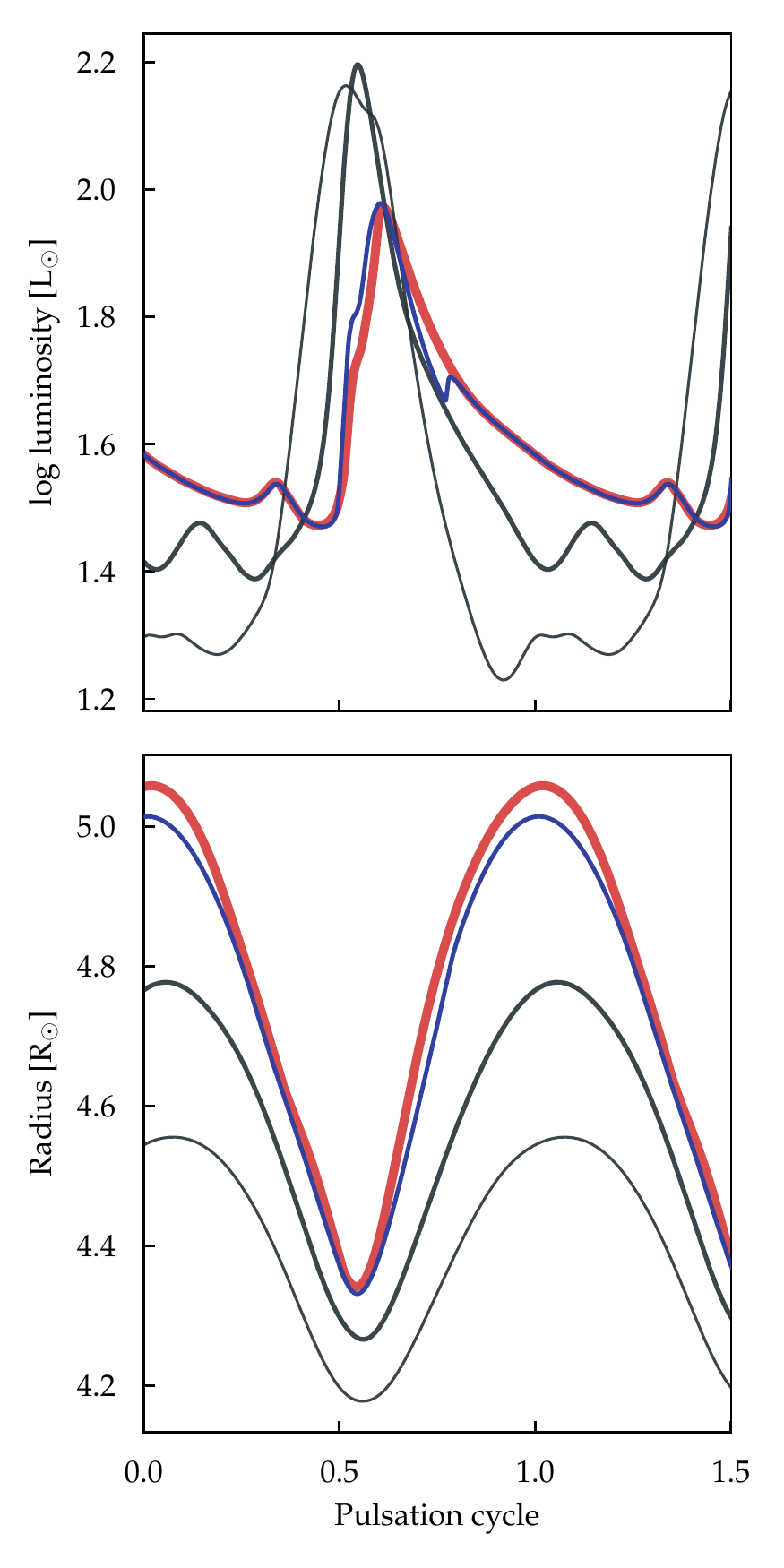}
\hspace{0.1cm}
\includegraphics[width=0.48\linewidth]{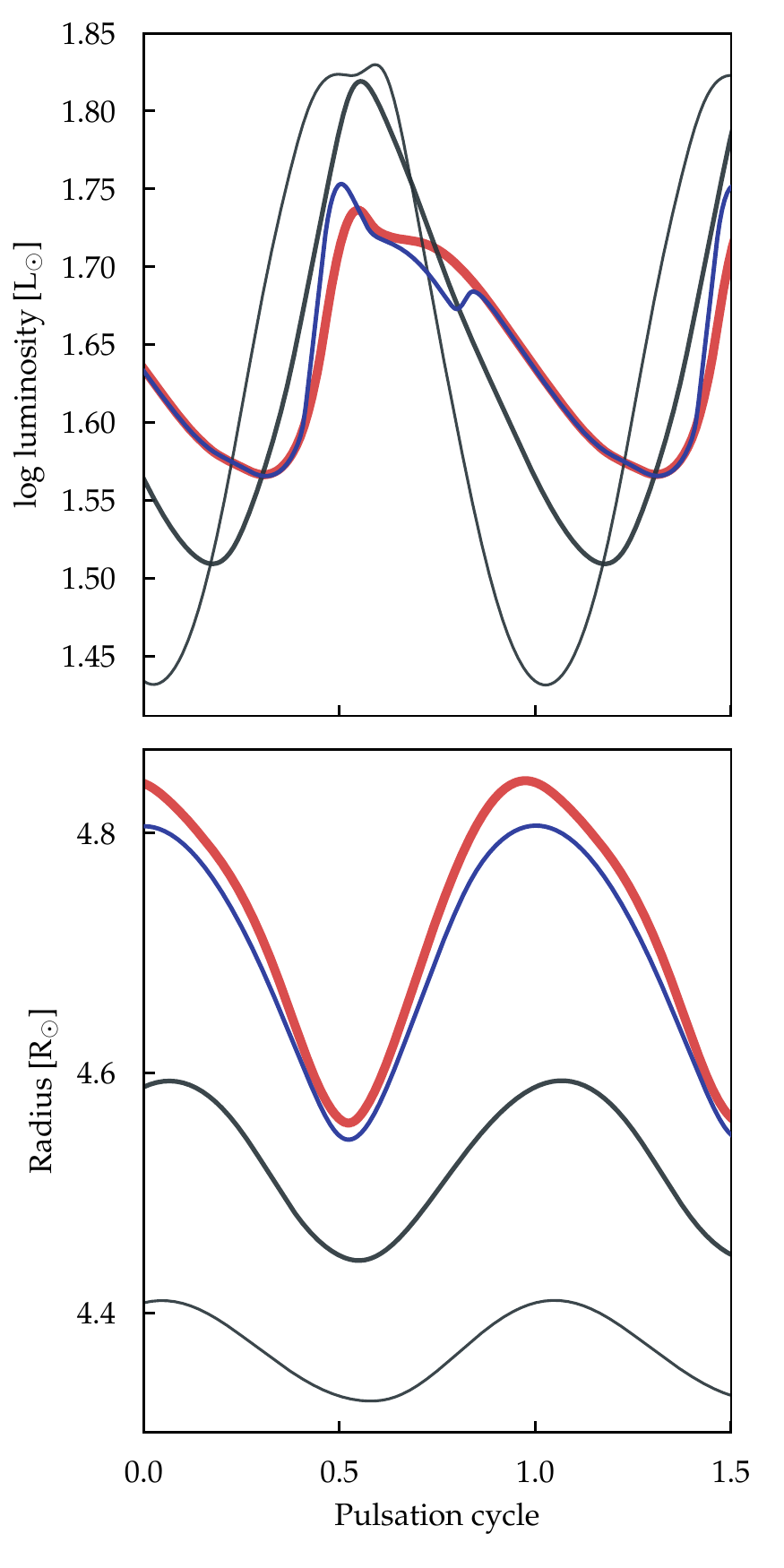}
\caption{Top: Comparison of the bolometric surface luminosity (red), 
	      the luminosity variation at the H/HeI (blue) 
	      and the HeII PIZ (cool edge heavier grey 
	      and hot edge thinner grey line)
          for RRab (left) and RRc model (right).
          Bottom: Respective radius variation; color coding is the
          same as in the top panels.
        }
\end{figure}

Resorting to the physical reasoning of \citet[][in Ch.~27.6 + 27.7]{pss68},  
in the linear quasi-adiabatic approximation applied to radiative regions 
it is found that $\delta L/L \sim \left( \Gamma_3 - 1 \right) \delta \rho/\!\rho$ and
$\sim \diff_x \left( \left( \Gamma_3 - 1 \right) \delta \rho /\! \rho 
\right)$.\sidenote[][-0.0cm]{
Notice:
	$\delta \rho / \rho = - 3\,\delta r / r 
	                                  - x\,\diff_x\left( \delta r / r \right)$
	with $\delta$ denoting a Lagrangian perturbation and $x\doteq r/\rast$.
	                                  }
Before a PIZ is encountered 
radiative dissipation ($\delta L/L$ rising outwards in Fig.~\ref{fig:RRLyrLINA}) 
grows with the density- and hence the radius-variation amplitude of the pulsation. 
Once a PIZ develops, $\Gamma_3 - 1$ drops and so does $\delta L/L$ accordingly.
Hence, the blocking of the energy flux at the bottom of the HeII PIZ depends 
on the time span over which the density is appropriately enhanced there.   
Figure~\ref{fig:RRLyr_LandR} illustrates the correlation of the  
luminosity variation as a function of depth in the stellar envelopes together with the
associated radius variations for RRab (left) and RRc (right) models, respectively. 
The top two panels show the bolometric luminosities emerging from the photosphere
(red line), the luminosity variation obtaining at the H/HeI-ionization zones 
(blue line), at the cool end of the HeII PIZ (thicker grey line), and finally 
at the hot edge of the  HeII PIZ  
(thinner grey line).
		\sidenote[][-2.0cm]{The luminosity variations of the RRab model 
		          yield:
		  \begin{tabular}[t]{lcc}\toprule
	  		Location     &  Sk &  Ac  \\ \midrule
			surface      & 5.7 &  2.0 \\ \addlinespace
			H/HeI PI     & 5.7 &  2.0 \\ \addlinespace 
			HeII PI, cool& 2.3 &  3.2 \\ \addlinespace
			HeII PI, hot & 2.0 &  2.2 \\ 
		  \bottomrule
		  \end{tabular}
		  
		 \medskip\noindent		
		 the RRc model yields:
		  \begin{tabular}[t]{lcc}\toprule
	  		Location     &  Sk &  Ac  \\ \midrule
			surface      & 3.2 &  1.1 \\ \addlinespace
			H/HeI PI     & 3.8 &  1.0 \\ \addlinespace 
			HeII PI, cool& 1.5 &  1.3 \\ \addlinespace
			HeII PI, hot & 0.6 &  1.0 \\ 
		  \bottomrule
		  \end{tabular}
		  \label{table:RRAc_Sk}		  		  
          }
The lower two panels contain the corresponding radius variations
at the same mass depths, using the same line scheme.
Evidently, for the RRab case on the left, the compression phase appears 
considerably narrower~--~looking more like a bounce~--~when compared with 
the sinusoidal lines belonging to both edges of the HeII PIZ. The emerging 
light curves pick up their rough form from what happens at the HeII PIZ. 
In case of the RRc model, the sharp
peak that emerges at the end of the ascending branch is contributed by 
the H/HeI PIZ. The reason of the more harmonic motion of the HeII PIZ in 
the RRc case can be attributed to the presence of the node at 
about $\log T = 5$. The proximity of the node inflicts a comparatively 
lower pulsation amplitude but a comparatively longer duration of compression
at the hot end of the HeII PIZ so that the nonlinearities are less expressed 
than in the unconfined F mode.  

Sidenote~\ref{table:RRAc_Sk} tabulates the variation of the quantities
Ac and Sk at various depths of the model envelopes. The skewness of the RRab
model doubles when going from the base of the HeII PIZ to the photosphere; the
variation of the acuteness, on the other hand, does not change significantly. In the
case of the F-mode pulsator, the action of the H/HeI PIZ steepens considerably the
ascending branch of the light curve. A similar behavior, with different numerical
values for the quantities Sk and Ac, is observed in the RRc model. Most 
importantly, the RRc's photospherical Sk value is much higher than what is observed 
in nature. This is caused by the sharp bump that is picked up across the 
H/HeI PIZ by the O1 pulsator. Because the bump that appears in our modeled case exceeds 
maximum light of the Grundform, the magnitude of Sk overshoots accordingly. 
If the maximum seen in the red-colored light curve in the
upper right of Fig.~\ref{fig:RRLyr_LandR} were weaker and would not count as the
global light maximum then the skewness would drop to a moderate Sk = 1.8, which compares 
favorably with what is observed say in the OGLE RRc case shown Fig.~\ref{fig:RRLyrObs}.

\begin{marginfigure}[-5.2cm]

\begin{center}
\includegraphics[width=0.99\linewidth]{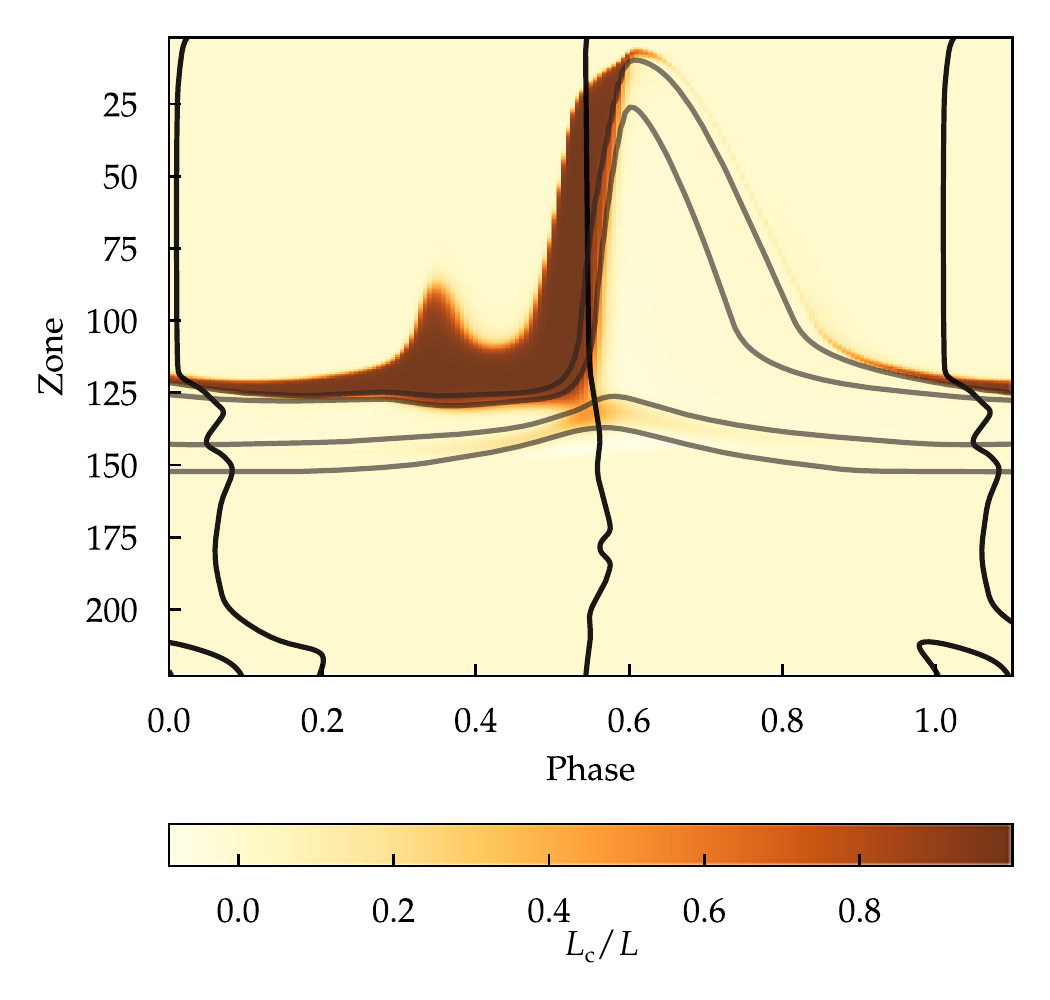}
\includegraphics[width=0.99\linewidth]{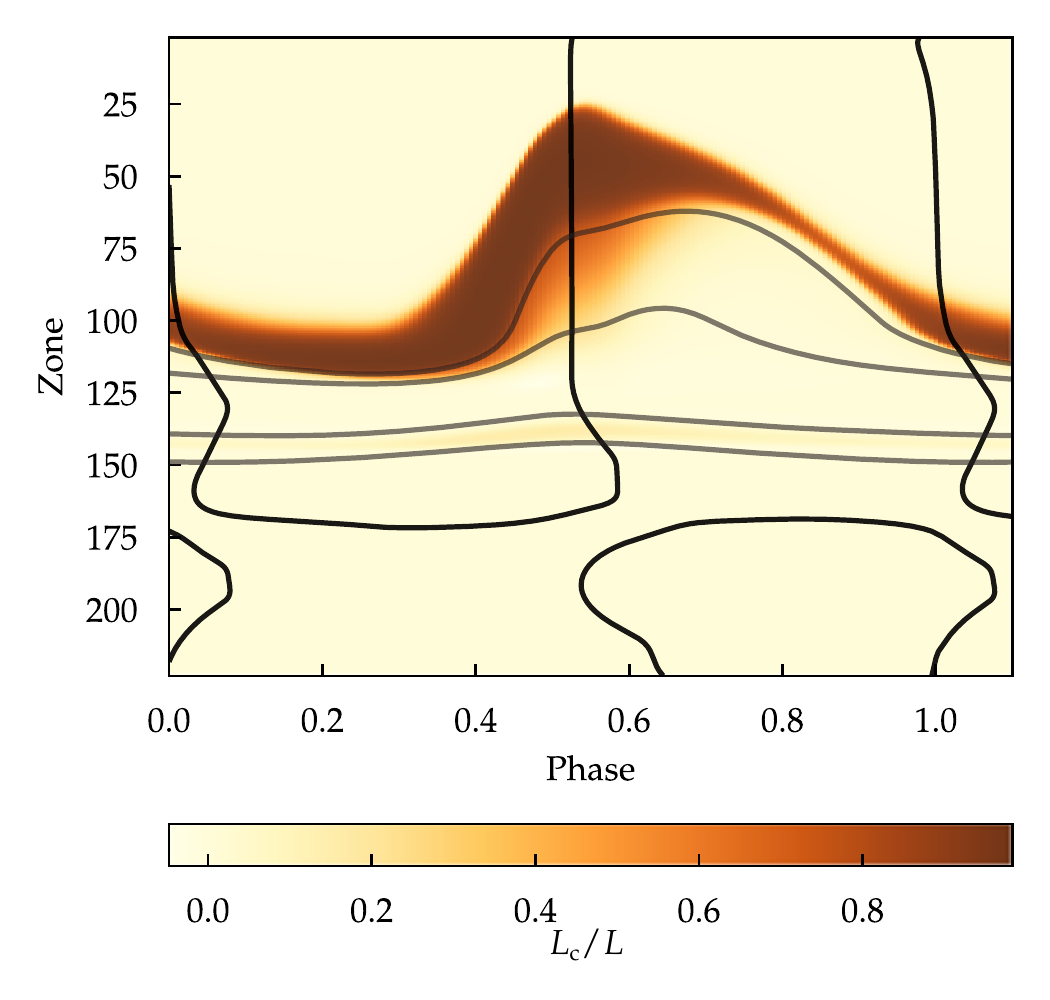}
\end{center}
\caption{Convective strength, measured as convective luminosity in units of 
the local total stellar luminosity as a function of depth in the envelope 
and of pulsation phase for the RRab (top) and the RRc model (bottom).}
\label{fig:RRLyr_convection}
\end{marginfigure}
For the RRab case, the quantities Ac and Sk of the model reproduce well what is
observed. RRc models reproduce the magnitude of the observed Ac values. The skewness, 
on the other hand depends strongly on the strength of a bump along the
ascending light branch the models pick up across the He/HeII PIZ. Across this
region, the star's energy is mainly transported by convection 
(see Fig.~\ref{fig:RRLyr_convection}). 
Strength and even position of such bumps are affected by the details of 
the time-dependent convection modeling, i.e. by the choice of the parameters entering
the \rsp\,computations. Since the convection-model parameters are usually 
weakly confined by theory, playing with them can be used to find better 
agreement with observations. 

Looking at the RRc triptychon of Fig.~\ref{fig:RRLyr_maps},
it is important to notice that the luminosity variation below (at lower mass) 
the 1O node is small or even 
negligible.\sidenote[][-2.2cm]{The same conclusion be drawn, of course, 
upon inspecting Fig.~\ref{fig:RRLyrLINA}} 
Hence, the effect of the phase change of the luminosity variation 
across the node is not pertinent for the pulsation. 

\section{Cepheid variables}

\begin{marginfigure}[-1.3cm] 
	\begin{center}
	\includegraphics[width=0.97\textwidth]{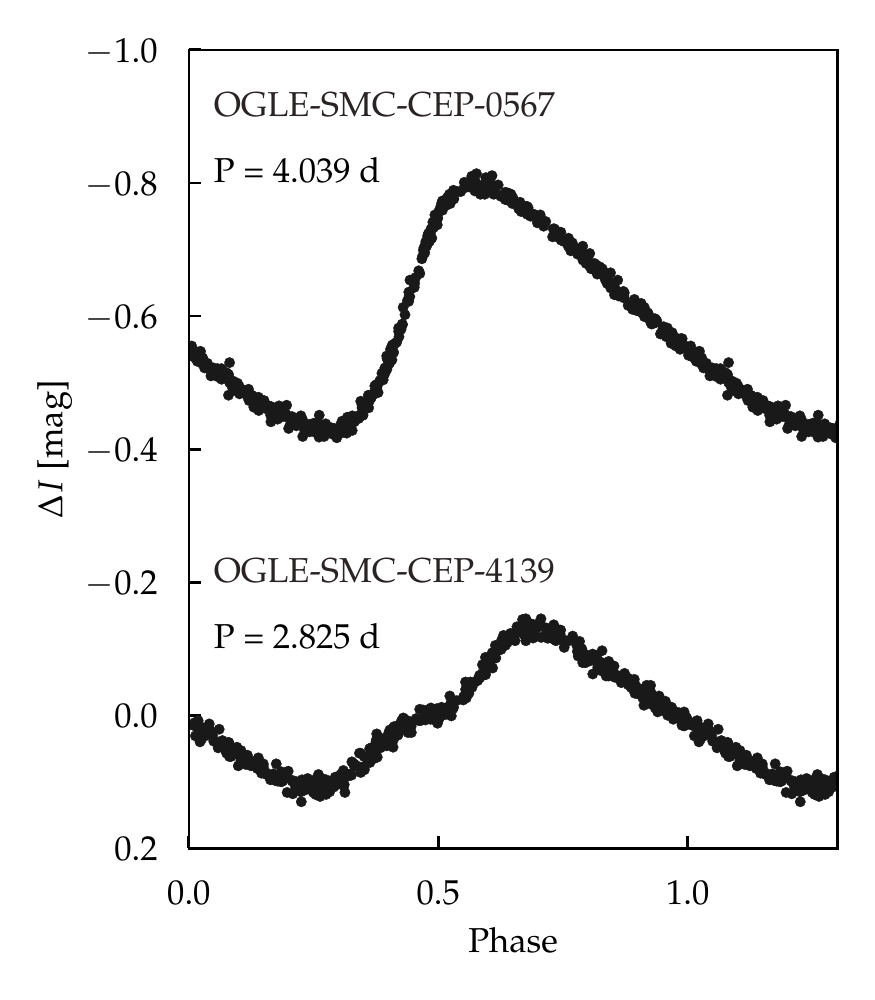} 
	\end{center}
     \caption{Light curves in the photometric $I$-band of Cepheid variables 
              observed in the SMC during the OGLE monitoring program.
              Top: F-mode with Sk = 3.0, Ac = 1.3. 
              Bottom: 1O-mode with Sk = 1.3, Ac = 1.0.  
             } \label{fig:CepObs}
\end{marginfigure}
The previous section addressed the differences in the luminosity variation
at the bottom of the helium PIZs of RRab and RRc pulsators. 
In this section we want to learn how the findings for RR Lyrae
variables carry over to the more massive, more luminous Cepheids. 
For this, we chose again an appropriate model star to describe a 
Cepheid whose fundamental \emph{and }first
overtone radial modes are both pulsationally unstable.
The stellar parameters of the Cepheid model are: $4.17\,\msol$ at 
$1439\,\lsol$ and $\teff = 6050$~K. The envelope is assumed to be chemically
homogeneous with SMC-compatible abundances of $X=0.75, Z=0.004$.
The Cepheid's envelope was subdivided into $300$ zones, with $80$
zones located at temperatures below $11\cdot 10^3$~K of the hydrostatic
initial model.

\begin{figure} 
	\begin{center}
	\includegraphics[width=0.88\textwidth]{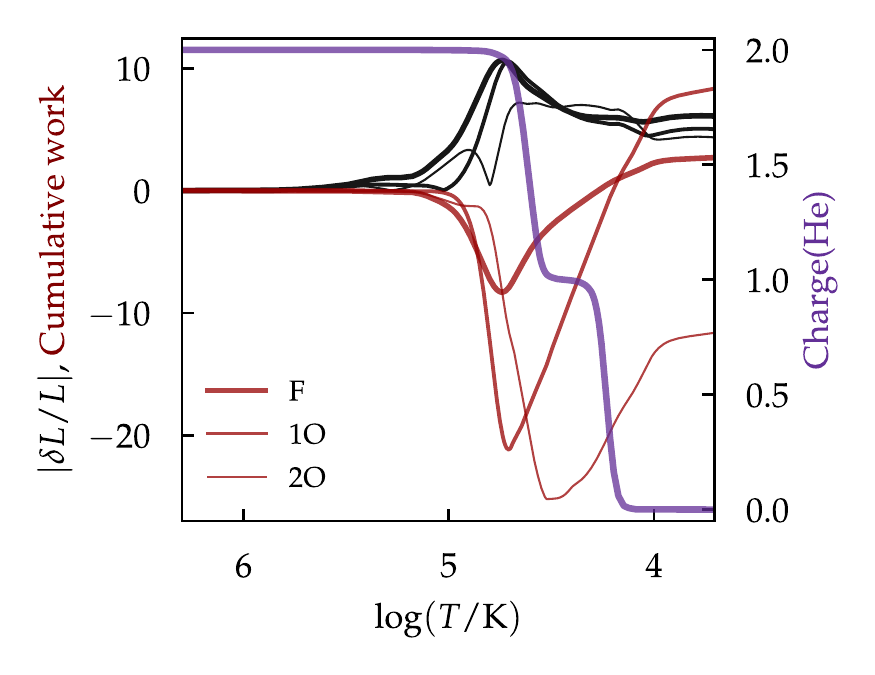} 
	\end{center}
     \caption{LNA results of the lowest three radial pulsation modes (coded
     	      by line thickness)
              of the Cepheid model. Temperature measures the depth
              in the model's envelope.
              The relative luminosity perturbation is plotted with black lines.
              The cumulative work done by the envelope is coded in red. 
              For orientation, the indigo line traces the average charge 
              of helium ions in the envelope.  
             } \label{fig:CepLINA}
\end{figure}

Figure~\ref{fig:CepLINA} depicts the LNA results for the lowest three radial modes 
of the short-period Cepheid model. The meaning
of the lines is the same as in Fig.~\ref{fig:RRLyrLINA}. As mentioned before, 
the F- and the 1O modes are unstable. The growth rate of the 1O-mode exceeds that
of the F-mode. As in the RR Lyrae case, the node of the 1O mode lies at about
$\log T = 5$ of the starting model.

\begin{figure*}
\label{fig:Cep_maps}
\centering
\includegraphics[width=0.48\linewidth]{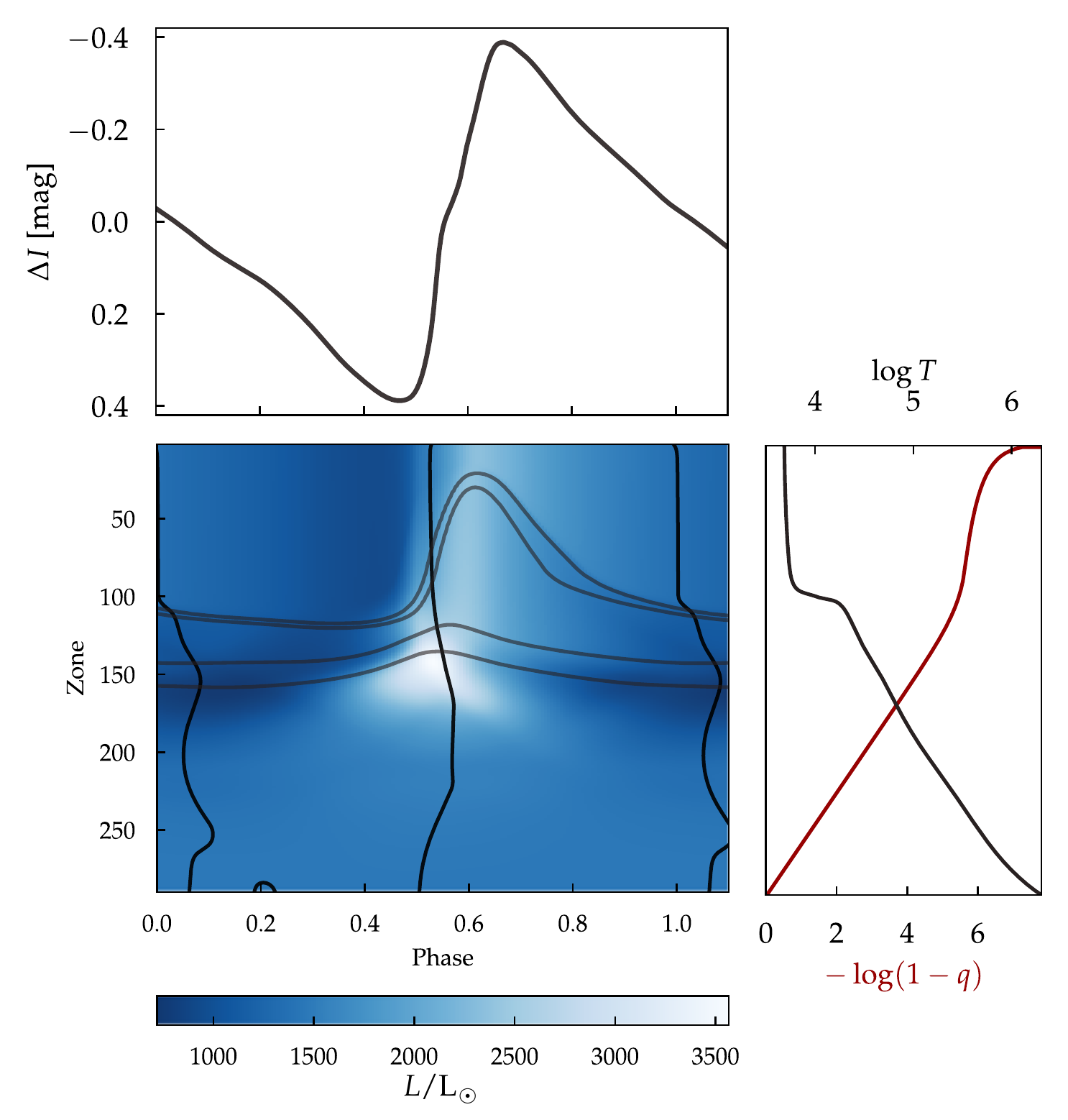}
\hspace{0.3cm}
\includegraphics[width=0.48\linewidth]{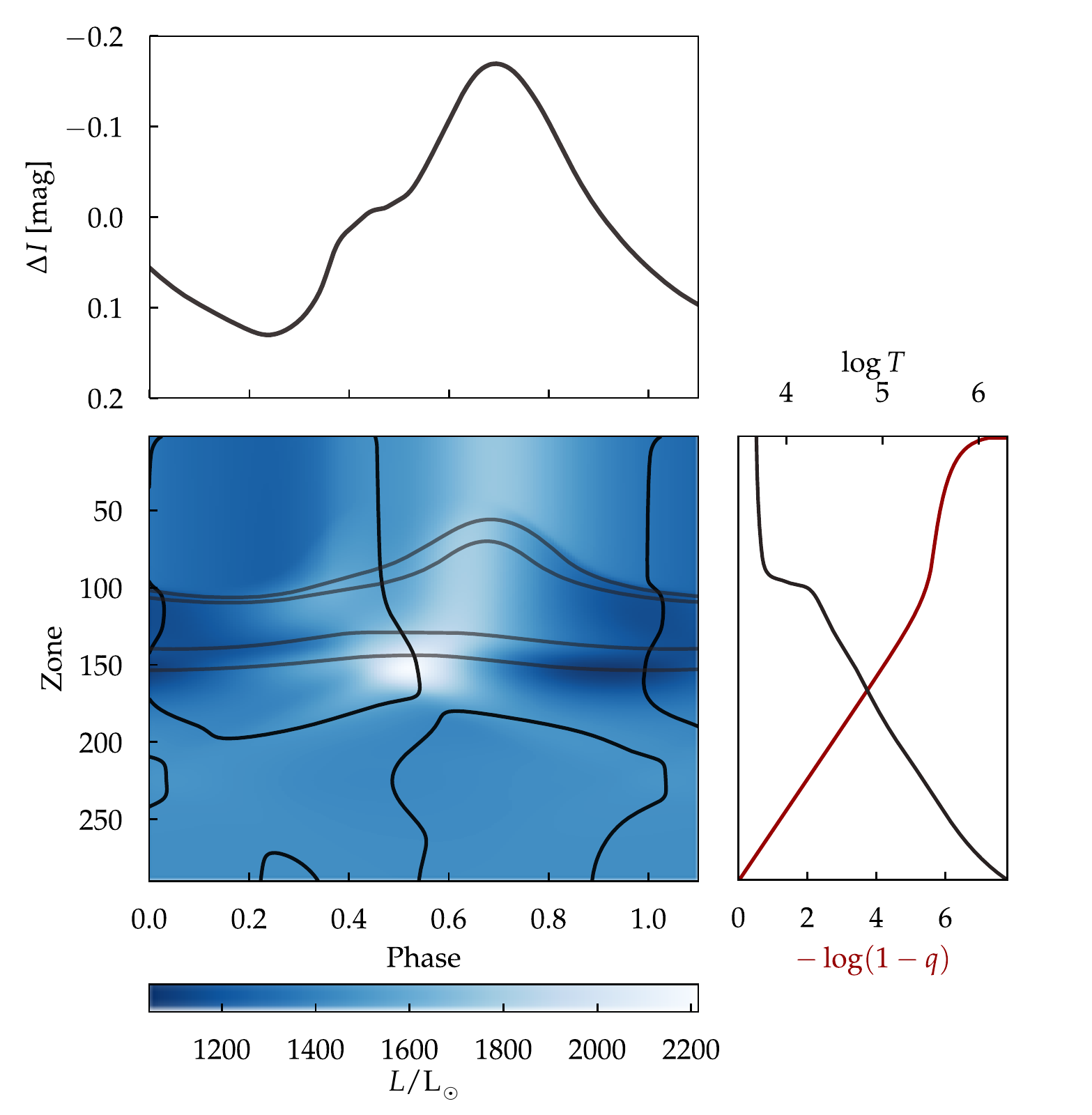}
\caption{Triptycha of the same kind as shown in Fig.~\ref{fig:RRLyr_maps} 
         for the F-mode with $P = 3.9325$~d (left) 
         and the 1O-mode with $P=2.83263$~d (right) Cepheid model.   
        }
\end{figure*}

The \rsp\,computations on the Cepheid model were started with pure 
F or 1O LNA velocity eigenfunctions whose amplitudes were set to $0.1$~km/s at 
the model surface. The F-mode pulsation took about 3500 cycles before 
it reached its limit-cycle with a period of $3.93925$~d, 
which agrees with the LNA period up to five digits. 
Even though the simulation was started with a pure LNA 
F-eigenmode the pulsations developed, between 1500 and 8000~d, 
a long mixed-mode phase (F and 1O-mode) with an upper envelope of the pulsation
amplitude at about $\Delta M_{\mathrm{bol}} = 0.4$ before the pure F-mode took 
over eventually and saturated at an amplitude of about $\Delta M_{\mathrm{bol}} = 1.1$.
The resulting light curve in the photometric $I$-band is shown in the top 
frame of the left triptychon in Fig.~\ref{fig:Cep_maps}. The quantities
plotted in the three frames are the same as in Fig.~\ref{fig:RRLyr_maps}. 
The F-mode's I-band Grundform parameters are Ac~=~1.1 and Sk~=~3.5, both of
which compare well with the respective observed quantities derived for the 
case shown in Fig.~\ref{fig:CepObs}. Also the luminosity behavior
displayed in the colormesh frame is comparable to the RRab situation. 

After perturbing the hydrostatic Cepheid model with the 1O velocity
eigenvector the pulsation grew exponentially into a pure 1O pulsation, 
which saturated at an amplitude of $\Delta M_{\mathrm{bol}} = 0.4$ 
after about 1000~days. The corresponding $I$-band light 
curve is displayed as the top frame of the right triptychon in 
Fig.~\ref{fig:Cep_maps}. The final nonlinear pulsation period
of $2.83263$~d is only 1~\permil\,longer than the corresponding LNA value. 
The magnitudes of the Grundform-characterizing quantities 
Sk~=~1.3 and Ac~=~1.6 fare decently well in comparison with the observed 
values for the 1O case shown in Fig.~\ref{fig:CepObs}. 
The bump at the lower ascending branch of the $I$-band light
curve, however, hints already at the impact its position and/or
amplitude can have on the magnitude of the quantity Ac. 
In the case of the 1O Cepheid the 
full width at half maximum to quantify Ac happens to lie just above the shoulder
of the ascending-branch bump. With a bump that occurred slightly higher, 
Ac would drop to unity or even lower. The observed 1O Cepheid, OGLE-SMC-CEP-4139, shows
also a bump after minimum light, its expression is, however, less pronounced
than that of the simulation. From the colormesh diagram
it is evident that the bump grows out of processes acting in the H/He PIZ 
so that details of this bump are prone to particular choices of 
convective parameters that characterize the coupling of the turbulent 
energy transport with the pulsating environment.  
The more or less horizontal components of the $v_{\mathrm{puls}} = 0$ 
curves in the colormesh diagram of the 1O mode trace the location 
of the pulsation mode's node. In contrast to the RRc case, the position of 
the $v_{\mathrm{puls}}$ node is not constant in zones, and hence 
\emph{not }constant in mass. The cyclic motion in mass is not large 
but noticeable anyway. 

\begin{figure}
\label{fig:Cep_LandR}
\centering
\includegraphics[width=0.48\linewidth]{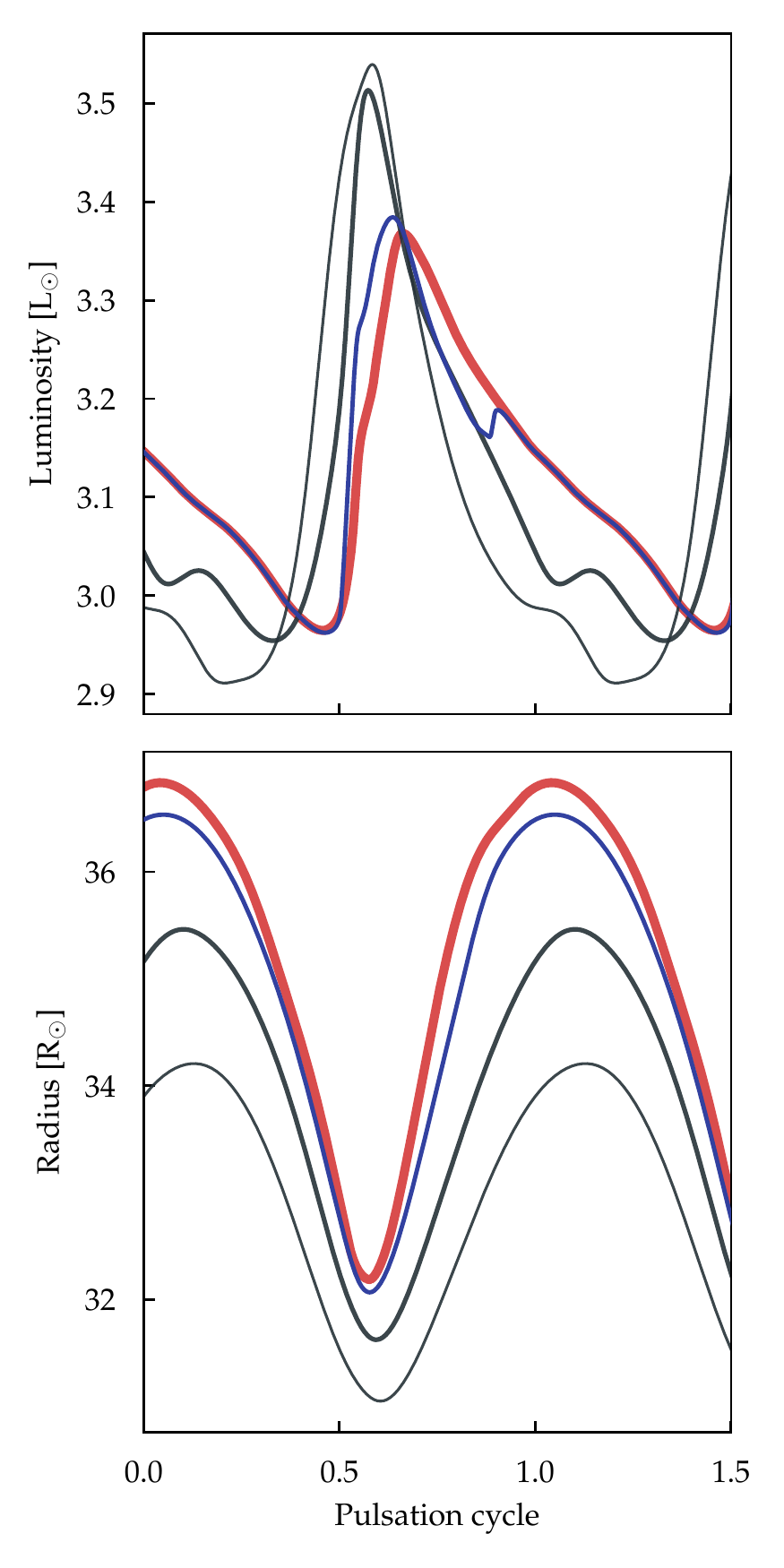}
\hspace{0.1cm}
\includegraphics[width=0.48\linewidth]{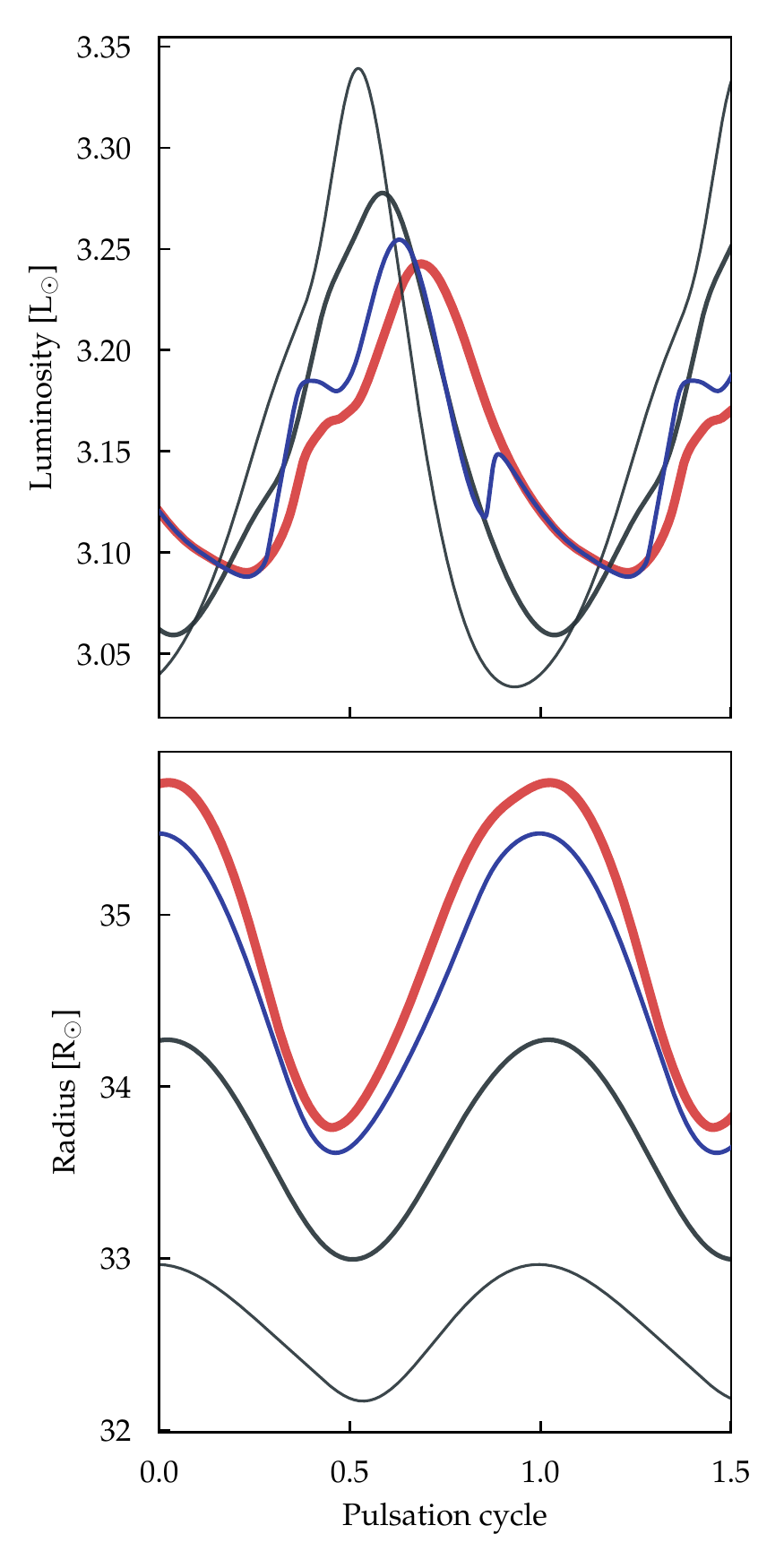}
\caption{Top: Comparison of the bolometric surface luminosity (red), 
	      the luminosity variation at the H/He (blue) 
	      and the HeII PIZ (heavier grey for the cool edge
	      and thinner grey line for the hot edge )
          for F- (left) and 1O-mode Cepheid (right).
          Bottom: Respective radius variation; color coding is the
          same as in the top panels.
        }
\end{figure}

\marginnote[-6.8cm]{
         \begin{tabular}[t]{lcc}        \toprule
	  		Mode   &  Sk    &  Ac    \\ \midrule
			F      &  3.5   &  1.1   \\ \addlinespace
			1O     &  1.3   &  1.6   \\ 
		                                \bottomrule
		  \end{tabular}         
}
Figure~\ref{fig:Cep_LandR} illustrates  how the luminosity and the radius 
varies at selected mass-depths through the envelope of the Cepheid model.
As in Fig.~\ref{fig:RRLyr_LandR} for the RR Lyrae model, the red line
traces the variation at the photosphere, the blue line at the H/He PIZ, 
the thick grey line at cool end and the thin line at the hot end of the HeII PIZ.
Comparing Fig.~\ref{fig:Cep_LandR} with Fig.~\ref{fig:RRLyr_LandR} reveals that the
Cepheid's F and 1O-mode behave much like those in the RR~Lyrae-variable model:
The amplitudes of the bolometric luminosity variation shrink with increasing mass.
The Grundform of the light curve is already carved out at the level of the 
HeII PIZ. Across the H/He PIZ, temporally well confined bumps and/or dips develop, 
which are eventually somewhat smeared out when they emerge at 
the photosphere. The radius variation of the F mode appears again more nonlinear 
with a sharp, bouncing episode around phase~$= 0.6$ than the smoother 
more sinusoidal-looking behavior of the 1O mode.

\section{Discussion}

The nonlinear hydrodynamical radial-pulsation instrument called \rsp\, in 
the recently released \mesa\, version \citep{paxton2019} allows to study in
detail fundamental and first overtone pulsations in RR Lyrae and Cepheid model
envelopes. Light curves of both pulsation modes could be reproduced satisfactorily. 

The Grundform of the light curves, which is quantified by the two parameters 
skewness, Sk, and acuteness, Ac,  can be attributed to the pulsation 
dynamics, i.e. to the duration of the compression phase, at the base of the HeII PIZ.
Already the duration of the imposed luminosity blocking determines the 
acuteness of the light curve. The Sk-affecting finer details such 
as bumps and dips along the light curve are influenced or even caused 
by processes in the H/He PIZ where superadiabatic convection dominates 
the energy transport in particular during the rising-light phase. 
Hence, tweaking convection-model parameters, which are theoretically ill 
constrained, can be used to mold these secondary features into the 
observationally documented shapes. The \rsp\,instrument
constitutes a versatile tool for such data-driven fine-tuning endeavors.

The compression phase of the F-mode pulsators is considerably more confined in time 
and nonlinear-looking than the more harmonic behavior encountered in 1O pulsators. 
The latter sinusoidal character is inflicted by the smaller pulsation 
amplitude at the base of the HeII PIZ due to the proximity of the 1O node at around
$10^5$~K. Hence, the node of the 1O mode is indeed the cause of the more
harmonic 1O light curves; however, it is not the swapped phase relation of the
energy flux from below the node as conjectured and modeled via OZMs in 
\citet{stellingwerf_overtone_1987} but the reduced pulsation amplitude in the 
neighborhood of the node. 

The $v_{\mathrm{puls}}=0$ lines in the colormesh frames of the triptycha of 
Figures~\ref{fig:RRLyr_maps} and \ref{fig:Cep_maps} indicate that maximum and 
minimum radius are not reached at the same phase for all mass depths. 
The phase of $v_{\mathrm{puls}}=0$ is not even a monotonous function of depth 
in the envelope. If the velocity nodal line of a 1O mode is added then the 
situation gets even more intricate: Pulsation phases can be found when up
to 6 locations in the envelope have $v_{\mathrm{puls}}=0$. Linear-theory thinking 
could mislead to interpret such snapshots as a manifestation of a 6O mode. 
Adhering, however, to a whole-cycle centered view makes it clear 
that one deals merely with an 1O mode. 
 
\newthought{Second Overtone Pulsations} are a natural conceptual next step 
to test the effect of the proximity to the HeII PIZ of the top-most node
on the Grundform of the light curve. The very existence of 2O modes in RR Lyrae variables 
in nature has been disputed in the past.
The latest compilation of the OGLE RR~Lyrae observations in the
Magellanic Clouds \citep{Soszynski2016} attributes to the RRc subclass 
all the variables that were previously suspected to be 2O pulsators. Furthermore,   
pulsation modelers never managed to find excited 2O modes for model 
stars that are compatible with evolutionary scenarios that explain 
single-star RR~Lyr populations \citep[e.g.][]{kovacs1998}. In accordance with the
older studies also our attempts failed
to obtain excited 2O modes over the whole domain that might contain them
according to Fig.~1 in \citet{Bono1994}. Therefore, we focused on 
Cepheids whose 2O modes are more established, observationally and theoretically.
Resorting to the \citet{Bono2001} study, we arbitrarily chose the case of a
$3.25\,\msol$ at $309\,\lsol$ with $\teff = 6850$~K star with and a 
homogeneous envelope composition of $X = 0.746, Z = 0.004$.

\begin{figure}
\label{fig:Cep_2O}
\centering
\includegraphics[width=0.7\linewidth]{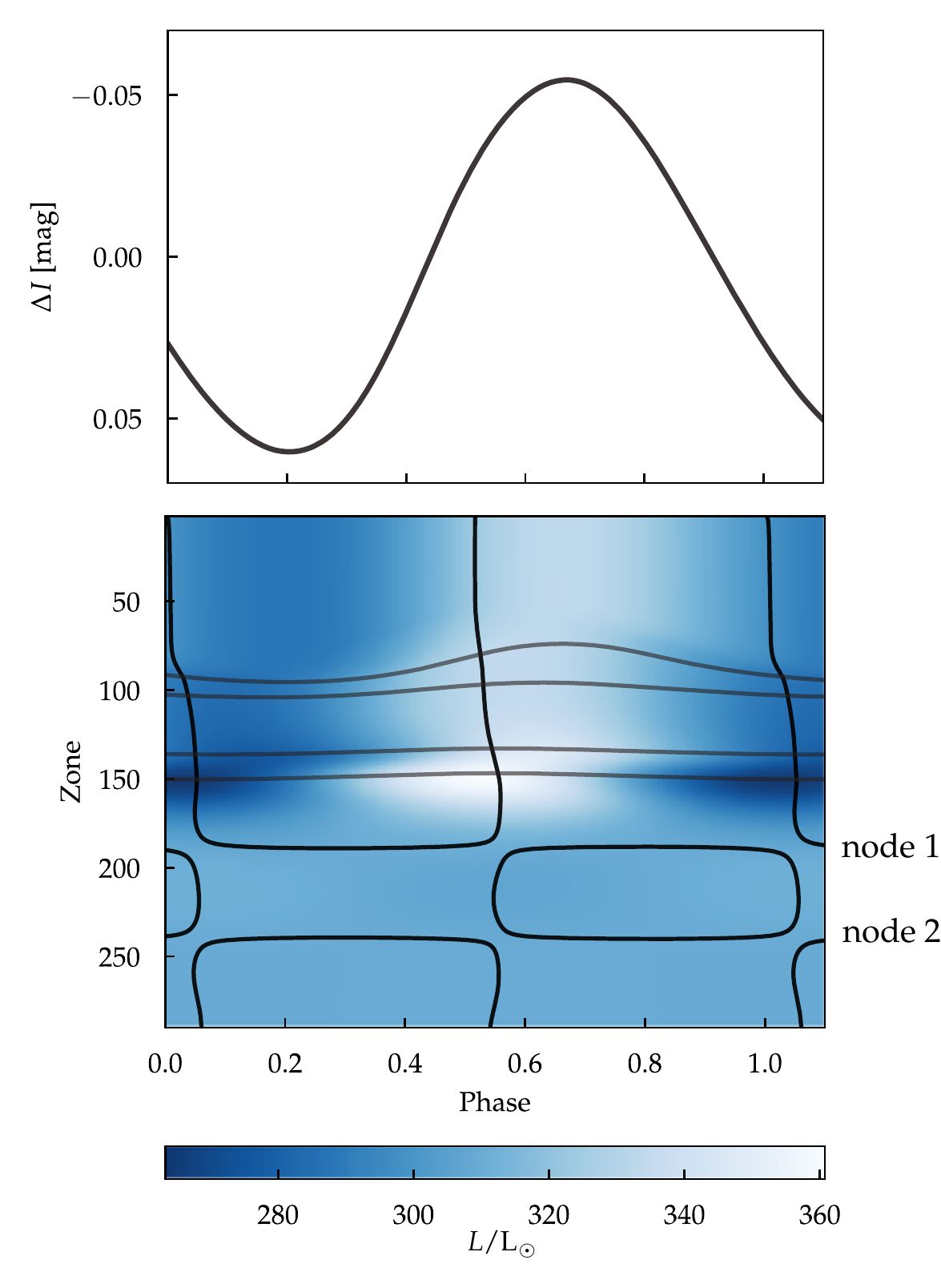}
\caption{ Second-overtone model Cepheid with $3.25 \msol$
		  after its pulsation saturated with a period of $0.49713$~d.
		  The top panel shows the I-band light curve with the same
		  phasing as the bolometric luminosity colormesh plot on the
		  bottom. The overplotted lines have the same meaning as in
		  Fig.~\ref{fig:RRLyr_maps}. 
        }
\end{figure}

The LNA analysis showed the 1O and 2O modes to be unstable with the 1O mode
only marginally so. After computing through about $10\,000$ cycles, 
the nonlinear pulsation was well saturated and essentially in its limit cycle with a 
period of $0.49713$~d. The resulting light curve in the photometric $I$-band is
shown in the top frame of Fig.~\ref{fig:Cep_2O} with its small amplitude
of about $\Delta I = 0.12$~mag. The bottom panel contains again
a colormesh plot with the color-coded bolometric luminosity as a function
of mass depth and pulsation phase. The black lines trace the 
loci of $v_{\mathrm{puls}}=0$; the grey lines mark the boundaries of 
the two helium PIZ as already described
for Fig.~\ref{fig:RRLyr_maps}.

The mass depth of the two nodes (referred to as node$\,1$ and node$\,2$ 
in Fig.~\ref{fig:Cep_2O}) of the 2O mode are hinted at
on the right side of the colormesh plot. In contrast to the 1O Cepheid case
discussed further up, here 
the essentially perfectly horizontal parts of $v_{\mathrm{puls}}=0$ indicate that
the mass depth of both nodes remains constant in time. Furthermore, 
node~1 lies closer (in mass) to the bottom of the HeII PIZ so that 
the light curve assumes an even more harmonic form than the 1O brethren. 
This is very much along the line of reasoning that the proximity of a node 
to the HeII PIZ sculpts the Grundform  of the light curve. As is evident 
from the colormesh plot, interior to node~1 and even more so to node~2 the 
luminosity variation is very small and is essentially negligible for 
the observable pulsation behavior.

\begin{figure}
\label{fig:Cep_FImpostor}
\centering
\includegraphics[width=0.7\linewidth]{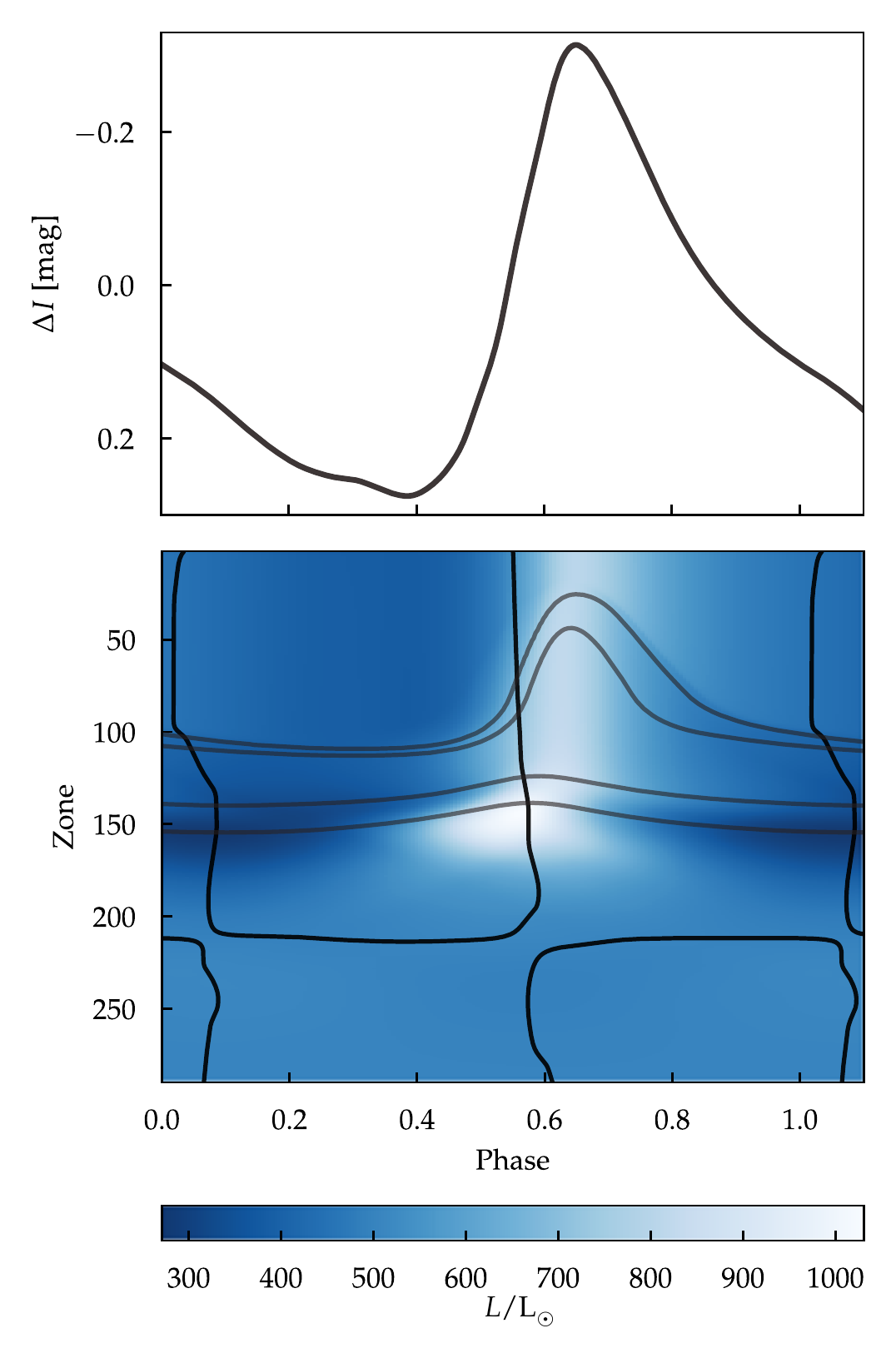}
\caption{Diptychon of the $3.8 \msol$ Cepheid F-mode impostor pulsating
with a period of $0.940$~d. The model pulsates in its 1O mode; except 
for the amplitude, the light curve, with Sk~=~3.0 and Ac~=~2.4, 
compares well with that of an F-mode pulsator.
        }
\end{figure}

\newthought{F-mode Impostors:} A model Cepheid with $3.8 \msol$ at $500 \lsol$ 
and with $\teff = 6600$~K was initiated with a 2O mode as obtained from the
linear stability analysis. During the first about $2000$ days
model time the mode evolved and saturated at about 
$\Delta\mathrm{M}_{\mathrm{bol}} = 0.2$~mag  amplitude. Since also the 
1O mode is unstable in this model, with about the same growth rate
as the 2O mode, some beating developed, which grew stronger 
over the ensuing roughly $3500$~days. Eventually, the pulsation of 
the stellar envelope settled at the considerably higher amplitude of 
$\Delta \mathrm{M}_{\mathrm{bol}} = 0.82$ and a period of $0.940$~d, which 
is very close to the LNA period of the 1O mode. This per se is not 
surprising, we just witnessed a mode switching from an initialized 2O mode 
to the apparently physically more attractive 1O mode. 
The real treat, however, is the final light curve of the 1O mode: 
The resulting light curve, shown in the top panel of Fig.~\ref{fig:Cep_FImpostor},
looks like that of an F-mode pulsator. In the $I$-band, Fig.~\ref{fig:Cep_FImpostor} 
reveals Sk = 3.0 and Ac = 2.4. The amplitude of the variability 
($\Delta I = 0.49$~mag) though is about $25 \%$ smaller than what 
a comparable generic F-mode model pulsator develops.

\begin{marginfigure}[-2.5cm] 
	\begin{center}
	\includegraphics[width=1.1\textwidth]{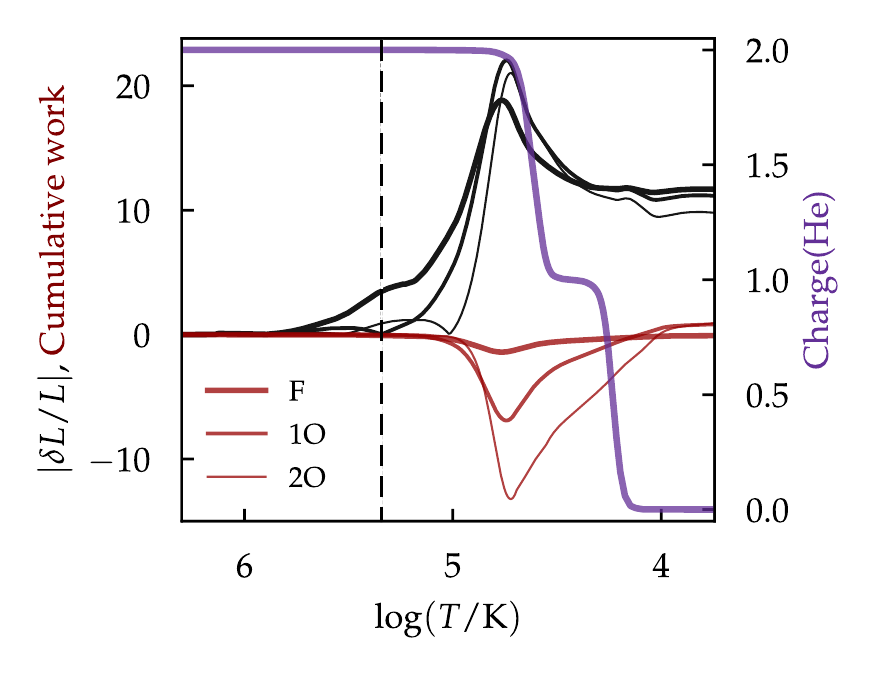} 
	\end{center}
     \caption{LNA eigendata for the $3.8 \msol$ Cepheid model. 
		     The vertical dashed line marks the position of the
		     luminosity node of the 1O mode at about $\log T = 5.34$.
		     The rest of the illustration is the same as in Fig.~\ref{fig:RRLyrLINA}.
             } \label{fig:Cep_FImpostorLINA}
\end{marginfigure}
Also in the F-mode impostor we see once again the effect of the location of the
pulsation-mode's node: The essentially horizontal black line at the level 
of about zone~210 in Figure~\ref{fig:Cep_FImpostor} traces the node of the velocity field. 
The 1O node in this low-luminosity Cepheid model lies deeper in the envelope 
than what was encountered in the other Cepheid and RR~Lyrae models (cf. 
Figs.~\ref{fig:RRLyrLINA}, \ref{fig:CepLINA} or 
\ref{fig:RRLyr_maps}, \ref{fig:Cep_maps}) for easier comparison we show also the LNA
results for the impostor in Fig.~\ref{fig:Cep_FImpostorLINA}; the vertical dashed line
indicates the position of the 1O node at about $\log T = 5.34$. Hence the 
\emph{lack of quenching} by the 1O node at the hot bottom of the HeII PIZ lets the
pulsation mode to develop an F-mode~--~like light curve. Light curves as 
described for our F-mode impostor are indeed observed in nature: Consult e.g. 
the short-period Cepheids plotted in the middle panel of Fig.~7 in
\citet{soszynski_optical_2010}. 
 
The F-mode impostors of the type discussed above for Cepheids have been 
encountered before also for RR~Lyrae variables. 
\citet{Bono1994} reported light curves of RRc models at low 
luminosity that mimicked RRab stars (see their Fig.~17). 
The authors pointed already then to the comparatively deep 
location in the envelope of the pulsation mode's node that 
favors the asymmetric light variability. Short-period 1O RR~Lyrae variables
that might be mistaken as F-mode pulsators are also very likely real: In the OGLE
database, cases can be found whose short periods would qualify them as RRc
variables but their light curves look distinctly RRab-like. Two 
representative examples are OGLE-LMC-RRLYR-09527 (P~$ = 0.258$~d, A$_I = 0.49$~mag, with
Sk~$ = 3.0$, Ac~$ = 1.4$) and OGLE-BLG-RRLYR-00767 (P~$= 0.321$~d, A$_I = 0.69$~mag, with 
Sk~$=4.2$, Ac~$ = 1.9$); more analogous examples can be found 
in the OGLE database with periods between 0.25 and 0.35~days.  
It should prove useful to find a method to discriminate between
real F-mode pulsators and F-mode impostors. In particular in the case
of RR Lyrae stars such a disentangling would allow to identify lower
luminosity RR Lyrae stars (the F-mode impostors) compared with RRc stars
of comparable period.   

It is important to keep in mind that the morphology of the 
light curves are not safe mode indicators: 
In principle at least, overtone pulsators can~--~photometrically~--~look like
F-mode pulsators. Browsing the light-curve catalogs of the OGLE project show
indeed that the breath of light-curve variations of 1O and 2O modes is considerable. 
This variation spectrum in nature is not necessarily associated with the
position of the outermost node alone, effects from convection-pulsation
interaction in the H/He partial PIZ likely inflict almost individually
shaped light variability on the respective pulsators.

\medskip

Browsing the OGLE~IV light-curve catalog for anomalous Cepheids (AC) reveals that
the unusually high percentage of about 60~\% of the 1O pulsators in the 
Magellanic Clouds have asymmetric $I$-band light curves and therefore a 
tendency to appear as F-mode impostors. If this latter character trait should
be uniquely tied to lower-than-usual $L/M$ ratios of the pulsators this might 
help to discriminate between competing scenarios of their origin. 

\begin{marginfigure} 
	\begin{center}
	\includegraphics[width=1.01\textwidth]{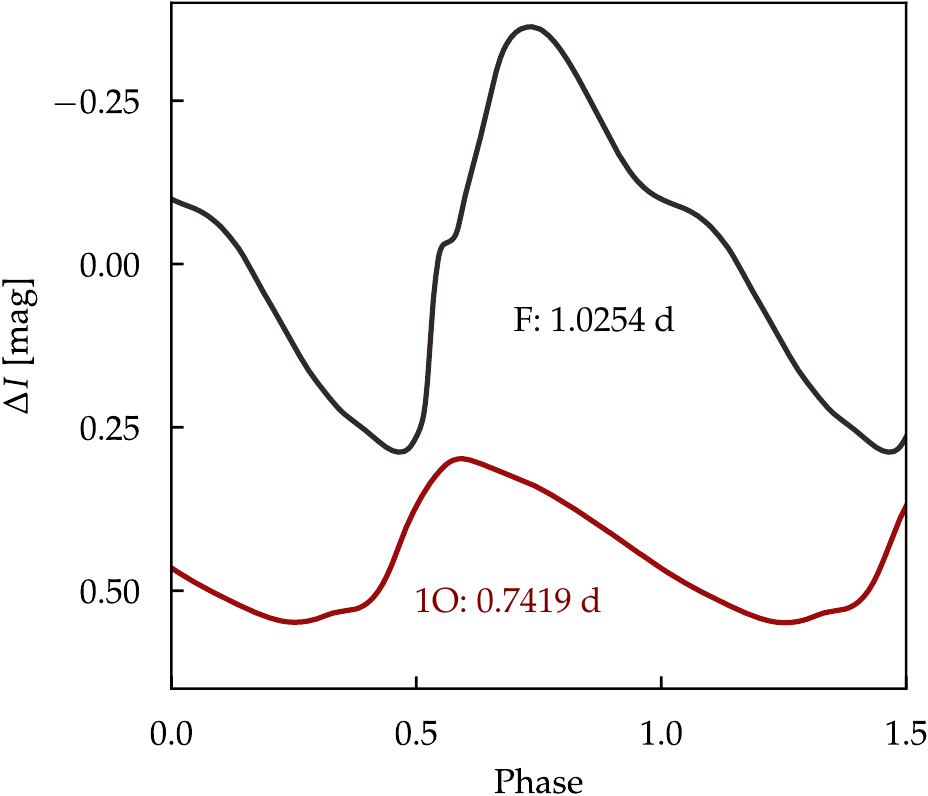} 
	\end{center}
     \caption{Exemplary $I$-band light curves the F- and 1O-mode  
     		  of a donor-turned-AC model
     	      with $0.68\,\msol$ at $100\,\llsol$, and $\teff = 6600$~K. 
		      The F-mode's I-band amplitude is $0.65$~mag, the one of the
		      1O mode $0.25$~mag. Keeping in mind that no fine-tuning was applied, 
		      these amplitudes compare decently well
		      with what is observed e.g. in OGLE~IV: $0.64 \pm  0.05$ for the
		      21 F-mode ACs with $1.0 \le \mathrm{P}/d \le 1.1$ and 
		      $0.37 \pm 0.03$ for the 14 1O-mode ACs with 
		      $0.7 \le \mathrm{P}/d \le 0.8$.    	      
             } \label{fig:ACs}
\end{marginfigure}
Preliminary \rsp\, computations have been performed on models fitting the 
binary scenarios as put forth in \citet{Gautschy2017} and on envelopes 
appropriate for the single-star scenario of \citet{Fiorentino2006}. 
Figure~\ref{fig:ACs} illustrates exemplarily light curves obtained for AC-like
envelopes. The particular case shown belongs to a model star of $0.68\,\msol$ at 
$100\,\llsol$, $\teff = 6600$~K, and with a chemical composition of 
$X=0.694, Z=0.006$ as appropriate for the binary-scenario of a 
donor-turned-AC case in \citet{Gautschy2017}. The fundamental mode has a period
of about one day, the (red colored) 1O mode has a cycle length of $0.742$~d. 
Asymmetric light curves, with Sk and Ac somewhat enhanced, of 1O modes were 
encountered frequently in all scenarios except for the merger model, which was 
also contemplated in \citet{Gautschy2017}. In accordance with the 1O's 
luminosity-perturbation node at a depth of around $10^5$~K, the more symmetrical 
merger-model light curves acquired low Ac and low Sk. In the respective period range, 
the computed I-band amplitudes of the 1O pulsations are on the lower side
compared with observations reported in the OGLE database.

Most of the AC-type F-mode light curves computed so far suffer from bulgy 
descending branches which fail to match observations. In particular, none of the
F-mode ACs in OGLE~IV with period between $1.0$ and $1.1$ days is observed
with a descending branch as computed and displayed in Fig.~\ref{fig:ACs}; 
all observed light curves in this period domain look like textbook cases 
of the RRab class with straight or even concave brightness declines.

\section{Conclusions}
Analyses of nonlinear hydrodynamical simulations of first-overtone 
RR Lyrae variables and of Cepheids revealed that the depth in the envelope 
of the respective node 
is instrumental to shape the Grundform of the emerging light curves.
In contrast to the conjecture  based on the OZM analyses of
\citet{stellingwerf_overtone_1987} it is \emph{not }the phasing of the 
energy flux across the node that determines the light-curve shape. 
We claim that it is the \emph{proximity }of the node to the hot 
edge of the HeII PIZ that sculpts the Grundform of the observable light curve.
Hence, it is essentially a \emph{quenching }effect of the node that 
inflicts a smaller-amplitude variability in the flux-blocking region, 
which then translates into a more sinusoidal, i.e. a harmonic light curve. 

Nonetheless, a 1O Cepheid model that developed a light curve mimicking 
an F-mode was encountered: In this case, the deep-lying node in the envelope 
can qualitatively explain the unusual light curve along our 
light-curve-shaping line of argumentation. Comparable behavior 
is also known to exist in some RRc cases. This experience alone 
should remind us again that light-curve morphology~--~for subclasses 
of pulsating stars overlapping in period and maybe even of comparable
amplitude~--~is not a safe classification tool.  

The preliminary pulsation computations of anomalous-Cepheid-like models
remain inconclusive on whether their light-curve behavior might help  
to discriminate between competing scenarios to explain their origin.

\medskip

\newthought{Acknowledgments}: This work relied heavily on NASA's Astrophysics
Data System. The OGLE light-curve atlas and 
database\sidenote{http:/\!/ogledb.astrouw.edu.pl/$\sim$ogle/OCVS/} 
served as an inspirational and quantitative source of information on 
pulsa\-ting stars in the sky. For this exposition, \citet{Soszynski2014} 
and \citet{Soszynski2015} were particularly relevant. Special thanks go 
to Hideyuki Saio whose pertinent comments helped to improve the manuscript. 

\newpage

\bibliographystyle{aa}
\bibliography{Overtones}        

\newpage

\section{Appendix}

Pertinent \texttt{inlist} settings used in the \rsp\,computations
(different zoning is mentioned in the main text) 
performed with \mesa\,version $11808$ were:

\begin{verbatim}
   ! controls for building the initial model
   RSP_nz       = 300   ! total number of zones in initial model
   RSP_nz_outer = 80    ! number of zones in outer region of initial model
   RSP_T_anchor = 11d3  ! approx temperature at base of outer region
   RSP_T_inner  = 2d6   ! T at inner boundary of initial model

   RSP_kick_vsurf_km_per_sec = 0.1d0 ! can be negative

   ! convection controls
   RSP_alfa   = 1.5d0   ! mixing length
   RSP_alfac  = 1.0d0   ! convective flux; Lc ~ RSP_alfac
   RSP_alfas  = 1.0d0   ! turbulent source; Lc ~ 1/ALFAS; PII ~ RSP_alfas
   RSP_alfad  = 1.0d0   ! turbulent dissipation; damp ~ RSP_alfad
   RSP_alfap  = 0.0d0   ! turbulent pressure; Pt ~ alfap     
   RSP_alfat  = 0.0d0   ! turbulent flux; Lt ~ RSP_alfat; overshooting.
   RSP_alfam  = 0.25d0  ! eddy viscosity; Chi & Eq ~ RSP_alfam
   RSP_gammar = 0.0d0   ! radiative losses; dampR ~ RSP_gammar             
\end{verbatim}
The rest of \rsp\, relevant \texttt{inlist} quantities were adopted as preset
in \texttt{inlist\_rps\_common}, which comes along with a \mesa\, installation. 

\end{document}